\documentstyle[amsfonts,12pt]{article}

\def\half{\textstyle{\frac{1}{2}}}

\def\A{{\frak A}}
\def\P{{\frak P}}
\def\ep{\epsilon}

\def\ra{\rightarrow}

\def\tint{{\textstyle\int}}

\def\s{\hskip.08em}
\def\d{\partial}

\def\ll{\lambda}

\def\b{\begin{eqnarray*}}     

\def\e{\end{eqnarray*}}       

\def\bn{\begin{eqnarray*}}     
  
\def\en{\end{eqnarray*}}       
  
\def\<{\langle}

\def\>{\rangle}

\def\no{\nonumber}

\def\{{\lbrace}

\def\}{\rbrace}

\def\vp{\varphi}
\begin{document}

\title{Overcoming Nonrenormalizability}
\author{John R. Klauder
\footnote{Electronic mail: klauder@phys.ufl.edu}\\
Departments of Physics and Mathematics\\
University of Florida\\
Gainesville, FL  32611}
\date{}     
\maketitle
\begin{abstract}
A suitable counterterm for a Euclidean space lattice version of $\vp^4_n$
theories, $n\ge 4$, is combined with several additional procedures
so that in the continuum limit the resultant quantum field theory is 
nontrivial. Arguments to support this unconventional choice are presented.
\end{abstract}
\section*{Introduction}
Perturbatively nonrenormalizable or renormalizable but not asymptotically
free quantized fields arise in theories of physical interest, such as in
quantum gravity or in the Higgs field of the standard model. As 
representatives of such fields we consider $\vp^4_n$ models for 
spacetime dimensions $n\ge4$. 

For $n\ge 5$, it is known that the quartically coupled relativistic 
scalar quantum field 
$\vp^4_n$ is  nonrenormalizable when formulated 
perturbatively, and trivial (equivalent to a free or
generalized free field) when constructed as the continuum limit of a 
Euclidean space lattice theory with a conventional lattice action \cite{fro}.
For $n=4$, the $\vp^4_4$ theory is perturbatively renormalizable,
not asymptotically free, and nontrivial despite strong evidence from lattice 
space computer studies that imply the theory is again trivial. Finally, we
observe that infinite-order perturbation theory also points toward triviality
due to the presence of Landau poles. It appears, therefore, that conventional

approaches lead to triviality.  In the present paper, we follow an 
unconventional path in our studies of $\vp^4_n$ theories, as we continue our 
quest \cite{kkk} to find nontrivial solutions for such theories.

Triviality may be readily characterized by the behavior of
selected lattice space correlation functions when the lattice 
spacing $a$ is very small. We assume units are chosen so that $a$ is a
dimensionless variable.
Mean field theory, generally regarded as valid when $n\ge 5$, leads to
\cite{fis} ($T$ denotes truncated)
  \bn   &&\hskip1.75cm\Sigma_k\s\<\vp_0\s\vp_k\>\propto a^{-2}\;,  \\
        &&\hskip-1cm\Sigma_{k_2,\dots,k_{2r}}\s\<\vp_0\s\vp_{k{_2}}
\cdots\vp_{k_{2r}}\s\>^T\propto a^{-2-6(r-1)}\;,  \\
        &&\hskip1.3cm\Sigma_k\s k^2\s\<\vp_0\s\vp_k\s\>\propto a^{-4}\;.  \en
Here, $\vp_k$ denotes the field value at the lattice site 
$k=(k_1,\ldots,k_n)$, $k_j\in\{0,\pm 1,\pm 2,\ldots\}$, $1\leq j\leq n$,
while $\vp_0$ is the field at the origin. To remove all 
dimensional and rescaling
aspects, as well as to allow for a continuum limit, one focuses on the 
quotients
   \bn  g_{r}\equiv -\s\frac{\Sigma_{k_2,\dots,k_{2r}}\s
\<\vp_0\s\vp_{k_2}\cdots
\vp_{k_{2r}}\s\>^T}
{[\Sigma_k\s\<\vp_0\s\vp_k\s\>]^r\,[\Sigma_k\s k^2\s\<\vp_0\s\vp_k\>/
6\s\Sigma_k\s\<\vp_0\s\vp_k\>]^{n(r-1)/2}} \propto a^{(r-1)(n-4)} \;.\en
When $n\ge 5$, $g_r\ra0$ as $a\ra0$ for all $r\ge2$ indicative for such
examples of a strictly trivial result. For $n=4$, mean field theory has
logarithmic corrections and it follows that 
   \bn  g_r \propto |\ln(a)|^{-(r-1)}\;,  \en
which still has the property that $g_r\ra0$ as $a\ra0$ for all $r\ge2$
leading again to the conclusion that the continuum theory is trivial. 

The dependence on $a$ that leads to the conclusion of triviality arises 
from the long 
range order that develops close to  a second-order phase transition. If, by
some procedure, we could simultaneously arrange that the {\it magnitude} 
of all correlation
functions was {\it uniformly rescaled for all $r\ge1$} so that 
  \bn  &&\<\s\vp_{k_1}\s\vp_{k_2}\s\cdots\s\vp_{k_{2r}}\s\>\propto a^{n-4},
\hskip1.79cm n\ge5\;, \no\\
  &&\<\s\vp_{k_1}\s\vp_{k_2}\s\cdots\s\vp_{k_{2r}}\s\>\propto 
|\ln(a)|^{-1},\hskip 1cm n=4\;,   \en
then
for all $n\ge4$ we learn that $g_r
\propto a^0=1$ for all $r\ge1$, and the door to nontriviality is open.
What follows is a conservative procedure to achieve the required
uniform rescaling that also promises to produce
a genuine covariant quantum field theory (after Wick rotation). 

\section*{Building to Specifications}
Our basic goal is to present a formula for the Euclidean space 
generating functional $S\{h\}$. To understand this formula,
it is helpful to construct it in a step-by-step procedure
from several different elements --
indeed, rather like a modern day ``powerpoint presentation''. To that end, 
we express
our proposal for the Euclidean space generating functional in the
schematic form
  \bn  S\{h\} =L_6\,P_5\,F_3\,N_2^{-1}\int \s e^{\s\Sigma h_k\s\vp_k\, a^n
\s-\s \A_1\s-\s{\frak P_4}}\;\Pi\,d\vp_k\;.  \en  
In this expression:\vskip.3cm
{\bf Element 1:} $\A_1$ denotes the conventional lattice action,
  \bn  \A_1\equiv \half\s\Sigma\s(\vp_{k^*}-\vp_k)^2\,a^{n-2}
+\half m_o^2(a) \s\Sigma
\s\vp^2_k\,a^n+\ll(a)\s\Sigma\s\vp^4_k\,a^n\;,  \en
where $k^*$ denotes each of the $n$ positive nearest neighbors to $k$,
the coupling constant $\ll(a)\ge0$, and 
the mass term $m^2_o(a)$ may even be
negative when $\ll(a)>0$. Sums extend over $k$ and $k^*$ as needed, and  
we suppose that the lattice is a large but finite hypercube with
periodic boundary conditions.\vskip.3cm
{\bf Element 2:} $N_2$ denotes a normalization factor,
  \bn  N_2\equiv\int\s e^{-\A_1}\;\Pi\, d\vp_k \;, \en
which implies that $D\equiv N_2^{-1}\s e^{-\A_1}$ is a probability density
distribution for the conventional lattice $\vp^4_n$ theory. 
 \vskip.3cm
{\bf Element 3:} $F_3$ is a dimensionless factor designed to uniformly 
rescale the conventional 
correlation functions, and is given by 
   \bn  &&F_3\equiv K\s a^{n-4},\hskip1.78cm n\ge5\;,  \no\\
      &&F_3\equiv K\s|\ln(a)|^{-1},\hskip1cm n=4\;,   \en
where $K$ is a fixed positive constant. We focus on
$F_3\le1$. Whenever $F_3<1$, the positive distribution 
$F_3N_2^{-1}\s e^{-\A_1}$ is no longer normalized. To fix the
normalization we introduce an additional term to the lattice action.
\vskip.3cm
{\bf Element 4:} ${\frak P_4}$ is an auxiliary and nonclassical 
($\propto \hbar^2$) factor added to the action which is given by
   \bn {\frak P_4}\equiv \half\, A(a)\s\Sigma\s[\vp^2_k-B(a)]/
[\vp^2_k+B(a)]^2\,a^n\;;  \en
why this particular form is chosen instead of a more familiar counterterm 
is discussed
below. At this point we only note that both $A(a)$ and $B(a)$ are 
positive, and that $B(a)\ra0$ as $a\ra0$. Therefore, 
the terms in $\frak P_4$ are significant for small field 
values ($\vp^2_k\simeq
B(a)$) which is where the terms in $\A_1$ are small; in contrast,
the terms in  $\frak P_4$ behave like $\half\s A(a)\s\vp^{-2}_k$
for large field values ($\vp^2_k\simeq a^{-(n-2)}$) which is where the 
terms in
$\A_1$ are 
important. The choice of $\vp_k^{-2}$ only involves the  dimensional 
factor $\hbar^2$, and requires no new dimensional coefficient for any $n$. 
As discussed below, there is a wide latitude available in which to 
{\it choose} the functional form of $B(a)$. Once that is done, however, 
the amplitude factor $A(a)$  
in $\frak P_4$ is then {\it determined} [in relation to the choice made 
for $B(a)$] by 
requiring that the distribution
   \bn  D'\equiv F_3\s N_2^{-1}\s e^{-\A_1\s-\s{\frak P_4}} \en
is {\it normalized} and hence is a probability density; an approximation 
for $A(a)$ [in relation to a given $B(a)$] is derived below. 
Furthermore, observe that
the correlation functions of $D'$ are similar to those of $F_3\s D$ since 
$\frak P_4$ has introduced changes
primarily for small $\vp$ values, which would tend to contribute 
relatively little to 
the correlation functions in the first place. 

We next need to restore the various correlation functions to 
macroscopic values. To that end we introduce multiple 
copies of the present system.\vskip.3cm
{\bf Element 5:} $P_5$ denotes a product over $N_R$ independent, identical
distributions, i.e., $P_5=\Pi_l$, $D'\ra D'_{l}$, $1\le l\le N_R$, all 
for the same test sequence
$\{h_k\}$ coupled to each factor. Here 
$N_R=[\!\![\s a^{-(n-4)}\s]\!\!]$ for $n\ge5$, while
$N_R=[\!\![\s|\ln(a)|\s]\!\!]$ for $n=4$, where $[\!\![\,\cdot\,]\!\!]$ 
denotes 
the integer part of its argument.  The resultant product has given the 
correlation functions macroscopic values as desired. 

With the specific construction as described above, the stage is set for:
\vskip.3cm
{\bf Element 6:} $L_6$ is the continuum limit $a\ra 0$ including, for 
convenience, 
a subsequent increase of the lattice volume in a natural way so that all of 
${\mathbb R}^n$ is covered. 

The resultant expression takes the form
  \bn  S\{h\}=\exp\bigg\{\s K\!\int[\s e^{\tint h(x)\s\vp(x)\,d^n\!x}-1\s]
\,d\sigma(\vp)\,\bigg\} \;,  \en
where $\sigma$ is a positive measure on fields that fulfills
$\tint\,d\sigma(\vp)=\infty$. As such we recognize $S\{h\}$ as determined
by a generalized Poisson process \cite{def}. The parameter $K$ may be chosen,
e.g., to satisfy some normalization condition on the two-point function.
\vskip.2cm{\hskip5cm*\hskip1cm *\hskip1cm *}

Elements 1 through 6 represent our recipe for resolving the ramifications
of triviality.\footnote{The author has long been concerned with
nonrenormalizable theories, and, in particular, counterterms that are
regularized versions of local inverse square field powers have 
previously been proposed \cite{new,kbook}. The version of ${\frak P}_4$ 
suggested in the
present paper involves only a single unknown function of the lattice
spacing, $A(a)$, and it is recognized here for the first time that 
normalization of $D'$ offers a clear requirement to uniquely determine
$A(a)$,
once we have made a choice for $B(a)$.
In previous work of the author it was hoped to restore the correlation 
functions to macroscopic values by an appropriate choice of the field
strength renormalization factor. In this paper it is recognized that this
suggestion is incorrect and it is replaced with the use of multiple copies. 
Of all
the author's many attempts to study nonrenormalizable theories, the
scenario offered in this paper seems to be the least arbitrary and most
compelling. It is also possible that similar procedures may have a wider
range of applicability.}

\section*{Frequently Asked Questions}
{\it Why have we chosen ${\frak P}_4$ as we have?}\vskip.1cm\noindent
Elsewhere (see Chapter 8 in \cite{kbook}) we have argued, for any
positive value of the coupling constant, that
nonrenormalizable interactions act as {\it hard cores in field space} 
projecting out certain fields that would otherwise have been allowed by the 
noninteracting theory alone. As such, the interacting theories are not even
continuously connected to the noninteracting theory as the coupling constant 
passes to zero! As a consequence, perturbation theory is a highly unreliable 
guide for what the counterterms to nonrenormalizable theories should be. The
choice made by ${\frak P}_4$ represents a local self interaction, without 
derivatives, which as a regularized form of an inverse square field
potential,
is arguably the only modification that can be introduced without 
qualitatively changing the original theory (c.f., 
${\bf P}^2=-\hbar^2\s\d^2/\d r^2+\hbar^2\s l(l+1)/r^2$ in quantum mechanics).

It is also the form that is suggested by certain idealized,
but fully soluble, nonrenormalizable quantum field models (see Chapter 10 
in \cite{kbook}). 

It is not too surprising that the choice of ${\frak P}_4$ has led to the fact
that $S\{h\}$ is expressed as a generalized Poisson process. What is novel
is that Elements 1 through 6 offer a relatively direct recipe to construct
the underlying measure $\sigma(\vp)$. 
\vskip.2cm
{\it How can modifying just the small field behavior change things?}
\vskip.1cm\noindent
While we have stressed that the introduction of ${\frak P}_4$ makes
relatively
small quantitative impact on the large-field behavior of the lattice
action, it must also be stressed that the small-field changes made by 
${\frak P}_4$ bring about a profound modification of the overall theory. 
This claim is already evident from the vast redistribution of the field 
probability in passing from the probability density distribution $D$ to
the probability density distribution $D'$.\vskip.2cm

{\it What does it mean to introduce multiple copies?}\vskip.1cm\noindent
The introduction of multiple copies implies that the underlying sharp-time 
field
operator algebra is {\it reducible}. A familiar example pertains to a
so-called generalized free field. It is clear that exploiting reducible 
representations opens up a new direction that is normally not pursued.
\vskip.2cm

{\it Does our procedure lead to a quantum theory?}\vskip.1cm\noindent
We first note that after Elements 1 and 2 one is dealing with the
conventional
Euclidean lattice space formulation of the $\vp^4_n$ problem. For a choice of
parameters which leads to a positive mass theory, it is natural to assume
that
the lattice expression will support a continuum limit that respects 
$n$-dimensional Euclidean invariance, reflection positivity, and 
clustering. Moreover, the factoring of 
the generating functional implicit in clustering should in no way be
effected by the addition of Elements 3 and 4, namely by rescaling the 
whole expression by $F_3$ and the introduction of another local potential
to restore normalization as represented by the term $\P_4$; in short, 
the long range correlations and
the associated decay of truncated correlation functions as two spacetime
regions become asymptotically separated should in no significant way be
altered
by the introduction of $\P_4$. The product of identical systems called
for in Element 5 does not interfere with invariance, reflection
positivity, or 
clustering. Consequently, assuming a uniform lower bound on a positive
mass, these important properties still ought to hold for the final
 result of our construction after the continuum limit. The relative
growth of the resultant correlation functions should also not be affected
by modifications at small field values. Therefore, although
these
arguments by no means constitute a proof, we do not anticipate
any conflict with requirements that our final theory admits a covariant
quantum field theory under Wick rotation \cite{gljaf}. $(\!\!($Indeed, if
coincident point singularities of the correlation functions are not
integrable, then our expressions for $S\{h\}$ should be understood as
standing
for a corresponding set of noncoincident point correlation functions.$)\!\!)$
\vskip.2cm

{\it Can one calculate something?}\vskip.1cm \noindent
Assuming that the continuum limit is relatively smooth, it is plausible that
certain aspects of the present model may
well be studied by the use of numerical simulation and Monte Carlo
techniques.
In particular, calculation of a nonzero value for $g_2$ as $a$ is made as
small as possible would support 
the expectation of a nontrivial continuum limit. \vskip.2cm

{\it Where should one begin?}\vskip.1cm\noindent
Observe that the suggested construction of $S\{h\}$ also permits us to
consider
the limit in which the coupling constant of the quartic interaction
goes to zero. Note well that the result of this 
limit is {\it not} the 
traditional free theory but is what is called the {\it pseudofree theory}. 
The pseudofree theory is worthy of examination in its own right. For
one thing, it is the theory to which the interacting theories are
continuously
connected (as opposed to the free theory). In addition, from a Monte Carlo 
point of view
the pseudofree theory is a natural place to begin since it deals with one
parameter less than the full theory. Clarification of the pseudofree theory 
would undoubtedly facilitate elucidation of the interacting theory.\vskip.2cm

Clearly the entire construction rests heavily on choosing a suitable function
$B(a)$, and then on finding the right function 
$A(a)$ to go with that choice. Let us next address that issue (in an 
approximate way, at least).

\section*{Choosing $B(a)$ and Approximating $A(a)$}
In this section we discuss how to choose $B(a)$ and develop an approximation 
for the all-important function
$A(a)$ that appears in ${\frak P}_4$, and in so doing we focus only on
Elements
1-4. As a preliminary to that analysis, 
however, we discuss a simpler, one-dimensional quantum problem as motivation.

The inverse square potential of interest to us can be illustrated in a 
simple quantum mechanical problem. In the Schr\"odinger representation, an 
eigenfunction of the form $\psi(x)=x^{-\gamma}\s f(x)$, with $f$ smooth, 
$\gamma>0$, and $x\ne0$, satisfies a Schr\"odinger equation of the form
  \bn -\s\frac{\hbar^2}{2\s m}\s\frac{d^2\s\psi(x)}{dx^2}+V(x)\s\psi(x)=0\;,  
\en
where the potential $V(x)$ takes the form
   \bn  V(x)\equiv\frac{\hbar^2}{2\s m}\s\frac{\psi''(x)}{\psi(x)}=
\frac{\hbar^2}{2\s m}\bigg[\s\frac{\gamma(\gamma+1)}{x^2}-2\s\frac{\gamma}{x}
\s\frac{f'(x)}{f(x)}+\frac{f''(x)}{f(x)}\s\bigg]\;. \en
If $f'/f\propto x$ near $x=0$, then the only singular term is $\hbar^2\gamma
(\gamma+1)/(2m\s x^2)$. Observe that this singular potential is invariant 
under $\gamma\ra-(\gamma+1)$; we only consider the form 
$x^{-\gamma}f(x)$ [rather than $x^{\gamma+1}f(x)$] since we are 
interested in eigenfunctions that have {\it enhanced} probability 
near $x=0$ due to the singular potential. 

Focusing on a small interval near $x=0$, let us assume for simplicity 
that $f=1$ in that region and then introduce a regularization for the 
singular eigenfunction that remains. To that end we change $\psi(x)$ 
to read $(x^2+\ep)^{-\gamma/2}$, $\ep>0$, and learn that 
 \bn -\s\frac{\hbar^2}{2\s m}\s\frac{d^2\s\psi(x)}{dx^2}+\frac{\hbar^2}{2\s
m}
\frac{\gamma(\gamma+1)\s[x^2-\ep(\gamma+1)^{-1}]}{(x^2+\ep)^2}\,\psi(x)=0\;, 
\en
which applies for all $x$ near to and including zero. For the regularized 
potential, it is noteworthy that the invariance under $\gamma\ra-(\gamma+1)$ 
is lost; in particular, the attractive well close to the origin evident in 
the regularized potential above, becomes repulsive when 
$\gamma\ra-(\gamma+1)$. The similarity of the one-dimensional 
regularized potential with the expression adopted in ${\frak P}_4$ is 
clear, and we next turn our attention to the field theory case. We 
predominantly treat the case $n\ge5$; comments regarding $n=4$ are made 
when appropriate.

The lattice action $\A_1+{\frak P}_4$ corresponds to a certain Schr\"odinger 
representation, lattice-space Hamiltonian operator as well. To develop a 
comparison with the one-dimensional example, let us first focus on the 
contribution of just the kinetic energy and the regularized, auxiliary 
potential in the lattice (Euclidean) action, which reads as ($k'$ refers 
to the eventual time direction)
 \bn  \half \Sigma\s(\vp_{k'}-\vp_k)^2\s a^{n-2}+\half\s A(a)\s \Sigma
\,[\vp_k^2-c\s B(a)]\s/\s[\vp_k^2+B(a)]^2\s a^n\;,  \en
where we have made one generalization to the form of the potential 
previously given. In particular, we have introduced a new (positive) 
parameter $c$; 
this change will let us see how $c=1$ is selected as the analysis proceeds. 
In the classical lattice Hamiltonian these two terms appear as
  \bn \half\s a^{-(n-1)}\Sigma\, \pi_k^2+\half\s A(a)\s \Sigma\,
[\vp_k^2-c\s B(a)]\s/\s[\vp_k^2+B(a)]^2\s a^{n-1}\;,  \en
where $\pi_k$ denotes the classical lattice momentum conjugate to the 
field $\vp_k$ at the site $k$. Correspondingly,
this part of the Hamiltonian operator becomes 
  \bn -\s\half\s\hbar^2\s a^{-(n-1)}\s\Sigma\,\d^2\!/\d\vp_k^2+\half\s 
A(a)\s a^{n-1}\s\Sigma\,[\vp_k^2-c\s B(a)]\s/\s[\vp_k^2+B(a)]^2\;.  \en
Now, in order that the auxiliary potential arises in the manner previously 
indicated for the one-dimensional example, it follows that
  \bn  && A(a)=\hbar^2\s a^{-(2n-2)}\s\gamma(\gamma+1)\;,  \\
       && \hskip.7cm c=(\gamma+1)^{-1}\;,   \en
where $\gamma>0$ remains undetermined. If $\gamma$ is a constant, this result 

suggests that $A(a)$ should diverge like $a^{-(2n-2)}$ as $a\ra0$, for any 
$n\ge4$. Below, however, we present arguments which suggest that although 
$A(a)$ may diverge as $a\ra0$, there is no compelling reason for $A(a)$ to 
diverge as fast as $a^{-(2n-2)}$. Accepting this argument implies that 
$\gamma$ is {\it not} a constant, but instead that
$\gamma=\gamma(a)$ and moreover that $\gamma(a)\ra0$ as $a\ra0$.
In fact, the property that $\gamma(a)\ra0$ is not too surprising when it 
is recognized that the normalization change introduced by $F_3$ is just 
$K\s a^{n-4}$, and that we have many field variables to contribute toward 
this relatively small change. If $\gamma(a)\ra0$ as $a\ra0$, then to leading 
order we may already set $\gamma+1=1$ and thus let $c=1$. (This possibility 
has already been anticipated earlier; see \cite{kbook}, p.~304.)

We now take up the question of estimating $\gamma(a)$, at least in a rough 
sort of fashion. For the normalization integral that enforces $D'$ to be a 
probability density, let us assume that the entire lattice volume 
${\cal V}\equiv(La)^n$ breaks up into an integral number $M$ of cells 
each of volume $v\equiv(\xi a)^n$, and that within each cell all the fields 
are completely correlated. On the other hand, we assume that fields in 
different cells are completely uncorrelated. With these strong assumptions, 
the normalization integral for $D'$ (for $n\ge5$, $\ll(a)=0$, $c=1$, and 
reverting to $\hbar=1$) takes the form given by 
\bn (K\s a^{n-4})^{-1/M}=\frac{\int \exp\{-\half v\s m_o^2\s\vp^2 -
\half v A(a)\s[\vp^2-B(a)]/[\vp^2+B(a)]^2\}\;d\vp}{\int\exp\{-\half
 v\s m_o^2\s\vp^2 \}\;d\vp}\;.  \en
A change of variables from $\vp$ to $\vp\s\sqrt{B(a)}$, along with the 
introduction of $E(a)\equiv v\s A(a)/2B(a)$, leads us to the relation
 \bn &&(K\s a^{n-4})^{-1/M}=2\sqrt{v\s m_o^2\s B(a)/2\pi}\\
&&\hskip.5cm\times\int_0^\infty \exp\{-\half v\s m_o^2\s\vp^2 B(a)-
E(a)\s[\vp^2-1]/[\vp^2+1]^2\}\;d\vp\;,  \en
which is an equation that implicitly defines $E(a)$ as a function of $a$ 
for sufficiently small $a$ where our real interest lies. For $n=4$, the 
left-hand side in this relation should be replaced by $(K/|\ln(a)|)^{-1/M}$. 
In such equations observe that we have isolated $v=(\xi a)^n$ -- the 
correlation volume -- which, near a second-order phase transition, is 
effectively a constant independent of $a$. 

On the basis of our approximate normalization condition, $E(a)$ is 
determined as a function of $a$, as well as a function of the other 
(constant) parameters, i.e., $K,\,v,\,m_o^2$, and $M$, along with the 
assumed choice for $B(a)$. An explicit expression for $E(a)$ is not 
possible, but we can at least seek an {\it approximate} expression for 
$E(a)$ which captures its {\it leading functional form for very small 
$a$}. This leading functional form is designed to capture the primary 
term $a^{-(n-4)/M}$ on the left-hand side as well as the form of $B(a)$, 
but it will not account for the constant factors: $K,\,v$, or $m_0^2$. 
Nevertheless, this leading dependence will be enough to suggest which 
form of $B(a)$ would be suitable. With that choice made we can then complete 
the estimate of $E(a)$ and hence determine the leading behavior of $A(a)$ 
in order to characterize our approximate model problem. All in all, this 
analysis will suggest a specific choice for $B(a)$ and offer an approximate 
form for $A(a)$, both of which can serve as starting approximations in a 
more careful analysis of the true normalization condition satisfied by the 
original expression $D'$.

Since $B(a)$ becomes small when $a$ becomes small, it follows that $E(a)$ 
must become large to maintain the normalization condition. This property 
may be seen
most easily if we rewrite the former integral in the form
  \bn  &&(K\s a^{n-4})^{-1/M} \\
&&\hskip-.5cm=2\sqrt{v\s m_o^2\s B(a)/2\pi}\int_0^1 \exp\{-\half v
\s m_o^2\s\vp^2 B(a)+
E(a)\s[1-\vp^2]/[1+\vp^2]^2\}\;d\vp \\
&&\hskip-.5cm + 2\sqrt{v\s m_o^2\s B(a)/2\pi}
\int_1^\infty \exp\{-\half v\s m_o^2\s\vp^2 B(a)-
E(a)\s[\vp^2-1]/[\vp^2+1]^2\}\;d\vp\;.  \en
Observe that the latter integral (from 1 to $\infty$) is bounded by one, 
so, for very small $a$, the former integral (from 0 to 1) must become 
large to accompany the large left-hand side. 
A steepest descent evaluation of the integral from 0 to 1 establishes that 
the leading behavior of this expression is given by
  \bn (K\s a^{n-4})^{-1/M}=\sqrt{v m_o^2B(a)/2\pi}\s\sqrt{\pi/3E(a)}\, 
e^{E(a)}\;.  \en
Taking the logarithm of both sides leads to 
  \bn  E(a)=[(n-4)/M]\,|\ln(a)|+\half\ln(E(a))+\half\s|\ln(B(a))|+O(1)\;, \en
where the latter term contains the parameters $K,\,v,\,m_o^2$, and other 
factors.

At this point in the analysis we need to choose $B(a)$. From a 
{\it mathematical} point of view, we could imagine choosing $B(a)$ as we 
like, e.g. $B(a)=a^2$, or even $B(a)=e^{-(1/a)}$. However, from a 
{\it computational} point of view, such choices of $B(a)$ are inappropriate 
since they tend to overwhelm the primary term of interest, namely 
$[(n-4)/M]|\ln(a)|$, especially when $M$ is large.  Therefore, to ensure 
that the primary term remains dominant we select (for $n\ge5$), e.g., 
   \bn  B(a)=|\ln(a)|^{-2}\;.   \en
With this choice, the equation for $E(a)$ becomes 
  \bn E(a)=[(n-4)/M]|\ln(a)|+\half\ln(E(a))+|\ln(|\ln(a)|)|+O(1)\;, \en
which has a leading order solution given by
  \bn E(a)=[(n-4)/M]|\ln(a)|+(3/2)\s|\ln(|\ln(a)|)|\;,\hskip1cm n\ge5\;. \en
{}From $E(a)$ we are led to
  \bn && A(a)=(2/v)B(a)\s E(a) \\
&&\hskip.96cm =2(n-4)\s(M\s v)^{-1}\s|\ln(a)|^{-1}+{\rm l.o.t.}\;, \\
  &&\hskip.96cm =2(n-4)\s(L\s a)^{-n}\s|\ln(a)|^{-1}+{\rm l.o.t.}\;,
\hskip1cm n\ge5\;,\en
where ``l.o.t." denotes ``lower order terms". There are several ways to 
interpret this result. On the one hand, we can imagine a 
limit (to be called limit I) in which $a\ra0$ and
where $L$ is held fixed which is rather the way a computer study might 
take place. On the other hand, we can imagine another limit 
(to be called limit II) in which, as $a\ra0$, we 
simultaneously increase $L$ so that the overall lattice volume 
${\cal V}=(L\s a)^n$ remains constant. Other limits are possible, 
but these two examples
illustrate the issue. If we choose limit I, then  
 $A(a)$ indeed diverges as $a\ra0$ but not as fast, for $n\ge5$,  as 
$a^{-(2n-2)}$. If we choose limit II, it follows as $a\ra0$ that $A(a)$ 
does {\it not} diverge,
but instead goes to zero. In either of these two limits, it follows, as 
described previously, that $\gamma(a)\ra0$ as $a\ra0$; specifically,
 \bn &&\gamma(a)=a^{(2n-2)}\s A(a)\\
&&\hskip.9cm =2(n-4)L^{-n}\s a^{(n-2)}\,|\ln(a)|^{-1}+{\rm l.o.t.}\;,
\hskip1cm n\ge5\;. \en
The reader can readily determine for themselves the vanishing behavior of 
$\gamma(a)$ as $a\ra0$ in either limit I or II.

For $n=4$ a slightly different argument applies. When $n=4$, the leading 
behavior of the basic equation becomes
  \bn (K/|\ln(a)|)^{-1/M}=\sqrt{v\s m_o^2\s B(a)/2\pi}\,\sqrt{\pi/3\s 
E(a)}\,e^{E(a)}\;, \en
which leads to 
 \bn E(a)=M^{-1}\s|\ln(|\ln(a)|)|+\half\ln(E(a))+\half|\ln(B(a))|+O(1)\;. \en
This time we choose 
  \bn B(a)=|\ln(|\ln(a)|)|^{-2}  \en
in order not to compete with the principal term. Therefore, we find that
  \bn E(a)=M^{-1}\s|\ln(|\ln(a)|)+\half\ln(E(a))+|\ln(|\ln(|\ln(a)|)|)|
+O(1)\;, \en
which has a leading order solution given by 
 \bn E(a) =M^{-1}\s|\ln(|\ln(a)|)|+(3/2)\s|\ln(|\ln(|\ln(a)|)|)|\;,
\hskip1cm n=4\,.\en
In turn, we find that
 \bn &&A(a)=(2/v)B(a)\s E(a) \\
&&\hskip.973cm =2\s(L\s a)^{-4}\s|\ln(|\ln(a)|)|^{-1}+{\rm l.o.t.}\;,
\hskip1cm n=4\;, \en
and
   \bn   &&\gamma(a)=a^6\s A(a)\\
 &&\hskip.89cm =2 \s L^{-4}\s a^2\s|\ln(|\ln(a)|)|^{-1}+{\rm l.o.t.}\;,
\hskip1cm n=4\;.\en
Once again, depending on the chosen limiting procedure (i.e., type I or II 
limits), it follows that $A(a)$ may diverge
or vanish, but even if $A(a)$ diverges that divergence is sufficiently 
slow to ensure that $\gamma(a)$ always vanishes as 
$a\ra0$.

The equations above tagged with $n\ge5$ and $n=4$ indicate, under the 
strong approximations made in this section, an acceptable choice of 
functions $B(a)$ and $A(a)$ [and thereby of $\gamma(a)$] whenever 
$n\ge5$ and $n=4$, respectively. In all cases, $\gamma(a)\ra0$ as
$a\ra0$, and as a consequence, {\it the change in the lattice action 
introduced, at each field point, actually vanishes as $a\ra0$}, relative 
to the kinetic energy contribution. Over the whole lattice, however, 
the {\it cumulative effect} of those soon-to-vanish individual contributions 
accounts for the desired change in the distribution. While the results 
obtained in this section surely depend on the strong assumptions that 
were introduced, it is nevertheless plausible that the results are 
sufficiently robust so that certain qualitative features of the solution 
survive even within a full calculation. We have in mind, that if we were 
to use the particular $B(a)$ developed in this section in the correct 
calculation of the normalization condition for $D'$, the resulting correct 
expression for $A(a)$ would still imply a correct form for $\gamma(a)$ 
that also vanished in the continuum limit. This conjecture is based on 
the fact that $\gamma(a)$ as derived in this section does not go to zero 
marginally, but, for all the cases discussed in this section and even in 
the least favorable limit (type I), $\gamma(a)$ goes to zero faster than 
$a^{2}$ as $a\ra0$, a fact which suggests that even if the functional 
form of $\gamma(a)$ changes, its asymptotic behavior, as $a\ra0$, may 
well be the same. Based on this assumption,
we have already set $c=1$ in the original version of ${\frak P}_4$.

In summary, the results of the analysis in this section suggest that 
suitable choices for $B(a)$ are given by
  \bn  && B(a)=|\ln(a)|^{-2}\;,\hskip1.95cm n\ge5\;, \\
   &&B(a)=|\ln(|\ln(a)|)|^{-2}\;,\hskip1cm n=4\;.  \en
It is also expected that these choices would prove satisfactory in a full 
analysis to determine the correct form of the remaining function $A(a)$ 
that enters the nonclassical, auxiliary
potential ${\frak P}_4$, and which should then render our version of the 
$\vp^4_n$ theory nontrivial for any $n\ge4$. As a place to begin to look 
for the
proper function $A(a)$ we can only recommend to start with the expressions 
found in the study in this section, namely,
   \bn  && A(a)= 2(n-4)(L\s a)^{-n}\s|\ln(a)|^{-1}\;, \hskip1.02cm 
n\ge5\;, \\
  &&A(a)=2(L\s a)^{-4}\s|\ln(|\ln(a)|)|^{-1}\;, \hskip1.38cm n=4\;.  \en
In these expressions, $L$ denotes the number of sites on each edge
of the lattice, while $(L\s a)^n$ denotes the lattice volume.

\section*{Acknowledgements}
Thanks are expressed to Bernhard Bodmann for his comments.

\end{document}